\documentclass[prb,showpacs,twocolumn,preprintnumbers,amsmath,amssymb]{revtex4-1}
\usepackage{amsmath}
\usepackage{amsfonts}
\usepackage{amssymb}
\usepackage{color}
\usepackage{graphicx}
\usepackage{bm}
\usepackage{ dsfont }
\newcommand{\beq}{\begin{equation}}
\newcommand{\eeq}{\end{equation}}
\newcommand{\beqa}{\begin{eqnarray}}
\newcommand{\eeqa}{\end{eqnarray}}

\def\to{\rightarrow}

\def\p{\partial}

\def\non{\nonumber }

\def\om{\omega}

\def\s{\sigma}

\begin{document}
\bibliographystyle{apsrev4-1}
\title{Gauge fields from strain in graphene}

\author{Fernando de Juan}
\affiliation{Materials Science Division, Lawrence Berkeley National Laboratories, Berkeley, CA
94720}
\affiliation{Department of Physics, University of California, Berkeley, CA 94720, USA}
\author{Juan L. Ma\~nes} 
\affiliation{Departamento de F\'isica de la Materia Condensada, Universidad del Pa\'is Vasco, 48080 Bilbao, Spain}
\author{Mar\'ia A. H. Vozmediano}
\affiliation{Instituto de Ciencia de Materiales de Madrid,\\
CSIC, Cantoblanco; 28049 Madrid, Spain.}
\date{\today}
\begin{abstract}
We revise the  tight  binding approach to strained or curved graphene 
in the presence of external probes such as Photoemission or Scanning Tunneling Microscopy experiments. We show that extra terms arise in the continuum limit of the tight binding Hamiltonian   which can not be accounted for by  changes in the hopping parameters due  to  lattice deformations, encoded in the parameter $\beta$. These  material independent extra couplings are of the
same order of magnitude as the standard ones and have a geometric origin. They include corrections to the position-dependent Fermi velocity and to a new vector field. We show that the new vector field does not couple to electrons like a standard gauge field and that no $\beta$-independent pseudomagnetic fields exist in strained graphene.

\end{abstract}
%
%
%
%
\maketitle

\section{Introduction}
One of the most interesting aspects of graphene is the tight relation between its morphological and
electronic properties. Although this issue has been explored at length in the theoretical literature
 \cite{NGetal09,VKG10}, and there is a fair amount of related experiments
\cite{BMetal09,TLetal09,DJetal10,HYetal10,YTetal10,TKetal11,GBetal11},   recent results
\cite{DWetal11,GMetal12,KZetal12} have given an extra push to the subject that will be explored in
this work.  

In the continuum limit of the standard tight binding (TB) approach, lattice deformations couple to
the electronic excitations in the form of  effective gauge fields and scalar potentials
\cite{SA02,VKG10}.  In particular, the so called pseudomagnetic fields have acquired a physical
reality after the observation
of Landau levels from strain in graphene samples \cite{LBetal10} predicted theoretically in
ref. \onlinecite{GKG10,GGetal10}.
These deformation gauge fields are at the basis of the proposed strain modifications of the 
electronic properties (strain engineering) of graphene \cite{GLZ08,PC09,GKG10,LGK11,JCetal11,CAH12}
and have been used in the design and modeling of recent experiments exploring the physics of lattice
systems either with cold atoms \cite{JBetal08,WGS08} or with artificial lattices made with CO
molecules in a Cu surface\cite{GMetal12}.

Hence it is very important to establish the accuracy and completeness of the TB description,
to ascertain if there are modifications to the model  and, if so, how they will  affect the experiments.

In the standard approach the parameter that links the TB electronics with the continuum elasticity theory,
$\beta$, is related to the electron--phonon coupling and appears in the definition of the
strain--induced  effective magnetic fields. $\beta$ reflects the changes in the hopping parameter $t$ of the TB model
with the changes of the relative distances between atomic nearest neighbors  due to the lattice
deformations.  
In a recent work \cite{KPetal12}
it was claimed that extra $\beta$-independent pseudomagnetic fields arise in the standard TB
description coming from the displacements of the atomic positions of the lattice. Following this work
there have been attempts to correct the previous calculations leading 
to ``strain Landau levels" \cite{ON12}. Moreover, inspired by a geometric approach to curved graphene
\cite{GGV92,GGV93,CV07a,CV07b,JCV07},  the continuum  TB Hamiltonian was supplemented  in
ref. \onlinecite{JSV12} with additional
$\beta$-dependent terms arising from a higher order derivative expansion,  which can be interpreted
as a position-dependent Fermi velocity and an new vector field.

In what follows we will show that   no  $\beta$-independent pseudomagnetic   fields exist  in
strained graphene: the only pseudogauge fields are the well-known $\beta$-{\it dependent} fields in
Eq.~\eqref{gaugefield}. We will indeed identify all the new terms arising from ``frame effects" 
(i.e., due to  the actual  atomic positions)  needed to complete the TB description whenever  the
system is coupled to external probes. But we will see that they only modify the coefficients of the
position-dependent Fermi velocity and new vector field obtained in ref. \onlinecite{JSV12}.  We will
further clarify the
nature of the new vector field and show that it does not act as a pseudogauge field,  although it
may have interesting physical effects, such as pseudospin precession.  We will also show  that the
extra gauge fields suggested in ref. \onlinecite{KPetal12} can be completely eliminated by a gauge
transformation and have no physical consequences. Finally
we will discuss the experimental context in which the newly derived  terms might lead to observable effects.

\section{Frame effects}

We will assume for simplicity that there are no short range interactions or disorder
connecting the two Fermi points of graphene, so that the low energy description around each point remains valid.
As  is well known in the  TB-elasticity approach \cite{GHD08,VKG10}, elastic deformations of the
lattice give rise, in the continuum limit, to vector potentials that  mimic the coupling of real  magnetic fields to the electronic current. The standard TB Hamiltonian  in the continuum limit is
\beq
H_{TB}=-iv_0\int d^2 x\psi^{\dagger}\sigma_j(\partial_j + i A_j)\psi.
\label{HTB}
\eeq
where $v_0=3\, t a/2  $ is the Fermi velocity for the perfect lattice, with $t$ the hopping
parameter for  nearest neighbors and $a$ the lattice constant; j=1,2 (summation over a repeated index is understood over
the article), and $\sigma_j$ are the Pauli matrices. The potential $A_j$ is related to
the strain tensor by 
\beq
A_1 = \frac{\beta }{2 a} (u_{xx}-u_{yy})\;\;,\;\;
A_2 = \frac{\beta }{2 a} (-2u_{xy}),
\label{gaugefield}
\eeq
where $\beta\!=\! \vert\partial \log t / \partial \log a\vert$. The strain
tensor is defined as $u_{ij} = \frac{1}{2} \left( \partial_i u_j + \partial_j u_i +\partial_i h
\partial_j
h \right)$, where $u_i$ and $h$ are in- and out-of-plane displacements respectively. Note that one usually assumes that crystal deformations are small and~\eqref{HTB} is valid only up to $O(u_{ij}^2)$ corrections. We will follow this practice for the rest of the paper.


As shown in ref. \onlinecite{JSV12},  if one uses the TB approach to go  one order  higher in the
derivative expansion, the Hamiltonian \eqref{HTB} becomes 
\begin{equation}
H_{TB} =  -i\int d^2 x\psi^{\dagger} [v_{ij}({ x}) \sigma_i \partial_j +  v_0 \sigma_i
\Gamma_i  + i v_0 \sigma_i A_i]\psi,
\label{HTBcomplete}
\end{equation}
where the field $A_i$ is the one given in  \eqref{gaugefield}, $v_{ij}$ is the tensorial and space
dependent Fermi velocity, 
$v_{ij} = v_0 \left[\eta_{ij} - {\frac{ \beta}{4}}(2 u_{ij} + \eta_{ij} u_{kk})\right],$
and $\Gamma_i$ is a  new vector field given by
\begin{equation}
\Gamma_i = \frac{1}{2v_0}\p_j v_{ij}=-\frac{\beta}{4} \left(\partial_j u_{ij}+ \frac{1}{2}\partial_i u_{jj}\right).
\label{Aprima}
\end{equation}

 The key observation of the present work is that  TB Hamiltonians describing strained graphene~\cite{GHD08,VKG10}, and~\eqref{HTBcomplete} in particular,  are commonly derived in  a specific reference system, the ``crystal frame". 
The reason is that the Bloch waves $a_k\!=\!\sum_x e^{-i\vec k\cdot\vec x}a_x$ used to diagonalize the TB hamiltonian  are written using the atomic \textit{equilibrium} positions
$\{ x\}$, which are regularly spaced and independent of  the crystal deformation. On the other hand, in the presence of strain the
positions measured in the ``lab frame" are the actual positions of the atoms $y_i$. The two sets of coordinates are  related by  $y_i=x_i+u_i(x)$, where $u_i$ is the in-plane horizontal displacement vector. Note that the vertical displacements $h$ are identical in both systems. In the classical theory of elasticity, crystal (lab) frame coordinates are usually referred to as Lagrangian (Eulerian) coordinates \cite{CL95}.

Thus, the TB hamiltonian~\eqref{HTBcomplete}  is actually the crystal frame hamiltonian $H_c(x)$. In order to describe the interaction of 
electrons with external probes or fields, we must use the lab frame hamiltonian $H_{Lab}(y)$, i.e., 
the TB Hamiltonian has to be rewritten in lab frame coordinates. The TB hamiltonian is the sum of the Dirac hamiltonian $H_0$ plus the terms induced by the lattice
deformations. As these are already $O(u_{ij})$, we have to compute  change-of-frame corrections 
only for the $u_{ij}$-independent piece $(H_{0})_c$ of the crystal hamiltonian. The computation is  simplified by using the symmetric convention for the derivatives of the fermion fields 
\beq
(H_{0})_c=-iv_0\int d^2 x\psi^{\dagger}_c(x)\s_i\overleftrightarrow{\p_i}\psi_c(x)
\label{Hcrystal}
\eeq
where  $\psi^{\dagger}\overleftrightarrow{\p_i}\psi\equiv
1/2(\psi^{\dagger}\partial_i\psi-(\partial_i\psi^{\dagger})\psi$ and the subscript in $\psi_c$ indicates that this is the fermion field operator in the crystal frame.
The derivatives transform according to
\begin{align}
\frac{\p}{\p x_i}=\frac{\p y_k}{\p x_i}\frac{\p}{\p
y_k} =(\delta_{ik}+\p_iu_k)\p_k =(\delta_{ik}+\tilde u_{ik}+\om\varepsilon_{ik})\p_k,
\label{1}
\end{align}
where $\tilde u_{ik}=(\p_iu_k+\p_k u_i)/2$ is the linear piece of the strain tensor and 
$\om\varepsilon_{ik}=(\p_i u_k-\p_k u_i)/2$. We also have to
transform the integration measure
\begin{align}
d^2x&=\left|\det\left( \frac{\p x_k}{\p
y_i}\right)\right|d^2y =
\left|\det(\delta_{ik}-\tilde u_{ik}-\om\varepsilon_{ik})\right|d^2y \nonumber \\
&\simeq(1-\tilde u_{ii}
)d^2y.
\label{2}
\end{align}
On the other hand, $\psi_c^\dagger \psi_c$ is the particle density operator in the crystal frame. As the number of fermions in any region should be frame independent, we must impose $\psi_c^\dagger \psi_c\, d^2 x = \psi^\dagger \psi\, d^2 y$, where $\psi(y)$ is the lab frame field operator. This implies
\begin{equation}
\psi_c(x)=\left|\det\left( \frac{\p x_k}{\p
y_i}\right)\right|^{-1/2} \psi(y),\label{rescaling}
\end{equation}
which exactly cancels the Jacobian in~\eqref{2}. The net result is
\begin{align}
-&i v_0\int d^2 x\psi^{\dagger}_c(x)\s_i\overleftrightarrow{\p_i}\psi_c(x)
\simeq \nonumber\\ -iv_0\int d^2 y & \left[ \psi^{\dagger}(y)\s_i
\overleftrightarrow{\p_i}\psi(y)+(\tilde u_{kl}+\om\varepsilon_{kl})(\psi^{\dagger}\s_k\overleftrightarrow{
\p_l}\psi) \right],
\label{trans}
\end{align}
where the derivatives in the last term act only on the fermion fields.  
Finally, the dependence on the antisymmetric piece $\om\varepsilon_{ij}$ may be eliminated by a
local rotation of the spinors 
\beq
\psi(y)\to e^{-\frac{i}{2}\om\s_3}\psi(y)\simeq\psi(y)-\frac{i}{2}\om\s_3\psi(y).
\label{3}
\eeq
Indeed, the identity $i\s_k\s_3=\varepsilon_{kl}\s_l$ shows that this rotation cancels the term proportional to $\om$ in~\eqref{trans}. A  contribution proportional to $\p_k\om$ vanishes as well due to the anticommutation relation $\{\s_3,\s_k\}=0$ for $k=1,2$. 
This  yields
\beq\label{main1}
H_{Lab}=H_{TB}+H_{Geom}
\eeq
where $H_{TB}$ is given by~\eqref{HTBcomplete} and
\begin{align}
H_{Geom}&=-iv_0\int d^2 x\,\tilde u_{kl}(\psi^{\dagger}\s_k\overleftrightarrow{
\p_l}\psi)\nonumber\\
&=-iv_0\int d^2 x\,\psi^{\dagger}\left[\tilde u_{kl}\s_k
\p_l +\frac{1}{2}(\p_l \tilde u_{kl})\s_k\right]\psi .
\label{main2}
\end{align}
In the last line we have used integration by parts to revert to the asymmetric derivative convention. 
Note that,  to first order in the strain, $\beta$-dependent terms are the same in both frames. 
Eqs.~\eqref{main1} and \eqref{main2} are the main results in this paper.

As $\beta\simeq 2$, 
the new $\beta$--independent terms in $H_{Geom}$ are of the same order of magnitude
as those  in the standard TB hamiltonian~\eqref{HTBcomplete}. In particular, the space--dependent Fermi velocity
derived in the TB formalism in ref. \onlinecite{JSV12} will become
\beq
v_{ij}=v_0 \left[\delta_{ij} - {\frac{ \beta}{4}}(2 u_{ij} + \delta_{ij} u_{kk})+\tilde u_{ij}\right]
\eeq
with the corresponding correction for the vector field
\beq 
\Gamma_i=\frac{1}{2v_0}\p_j v_{ij}=-\frac{\beta}{4} \left(\partial_j u_{ij}+ 
\frac{1}{2}\partial_i u_{jj}\right)+\frac{1}{2}\p_j \tilde u_{ij}.
\eeq


The hamiltonian $H_{Lab}$ can also be obtained by performing the TB calculation directly in the lab
frame. This derivation is explicitly given   in  the Supplemental Material, where we also show that
the additional pseudogauge field found in ref. \onlinecite{KPetal12} has zero curl  everywhere and
can be eliminated  by a gauge transformation of the electronic wave function.
As $\Gamma_i$ is the only ``new" vector field in strained or curved graphene, 
in what follows we will comment briefly on its physical significance and compare 
it with the well known  pseudogauge field $A_i$ in Eq.~\eqref{gaugefield}.
First of all, note that, unlike $A_i$, $\Gamma_i$ is not a functionally independent field. 
The reason is that  the hamiltonian~\eqref{HTBcomplete} is hermitian only 
for $\Gamma_i \!=\! \frac{1}{2v_0}\p_j v_{ij}$. Thus,  a position 
dependent Fermi velocity requires the existence of the new vector field $\Gamma_i$.

A look at~\eqref{HTBcomplete}  might suggest that $\Gamma_i$ is some sort of purely 
imaginary~\footnote{As a consequence of the extra $i$, the vector field $\Gamma_i$ is odd under time reversal and couples with equal signs at the two Fermi points.} counterpart to $A_i$.  However, this is obviously wrong, as 
gauge potentials have to be real (hermitian).  The true nature of $\Gamma_i$ 
is  made apparent  if we use the identity $i\s_k\s_3=\varepsilon_{kl}\s_l$ to rewrite the relevant term as 
$-i v_0\s_i\Gamma_i = v_0\s_i \tilde\Gamma_i $, with
\begin{equation}\label{redef}
\tilde \Gamma_1 =  \Gamma_2 \s_3\;\; , \;\; \tilde \Gamma_2 =  -\Gamma_1 \s_3 .
\end{equation}
Note that  $\tilde \Gamma_i$ is matrix-valued and hermitian. This shows that the vector field $\Gamma_i$ plays the role  a hermitian connection for the  $SO(2)$ group of local pseudospin rotations~\eqref{3} generated by $\s_3$.
As a consequence, a position dependent Fermi velocity will  be accompanied by pseudospin rotation
(``pseudospin precession"), i.e., by electronic transitions between the two sublattices. In more
physical terms, whereas electrons propagating in a (pseudo)gauge field acquire a path-dependent
complex phase, the new vector field induces pseudospin rotation, very much like an optically active
medium turns the polarization plane of light. Thus $\Gamma_i$ is not a gauge field 
and  can not  give rise  to the characteristic  Landau levels of real or pseudo-magnetic fields: 
the only  pseudogauge field  in strained graphene is the well known $A_i$ given by~\eqref{gaugefield}.
Note also that,
 in general,  observable effects will not be associated to  the field $\tilde \Gamma_i$ itself but to
its curl, which by \eqref{redef} is proportional to the divergence of $\Gamma_i$. This is even more
obvious in the covariant model\cite{JSV12}, where $\Gamma_i$ appears as the spin connection
associated to fermions propagating in a curved background and its divergence is proportional to the
scalar curvature $R$. 
\begin{figure}[h]
\begin{center}
\includegraphics[width=8cm]{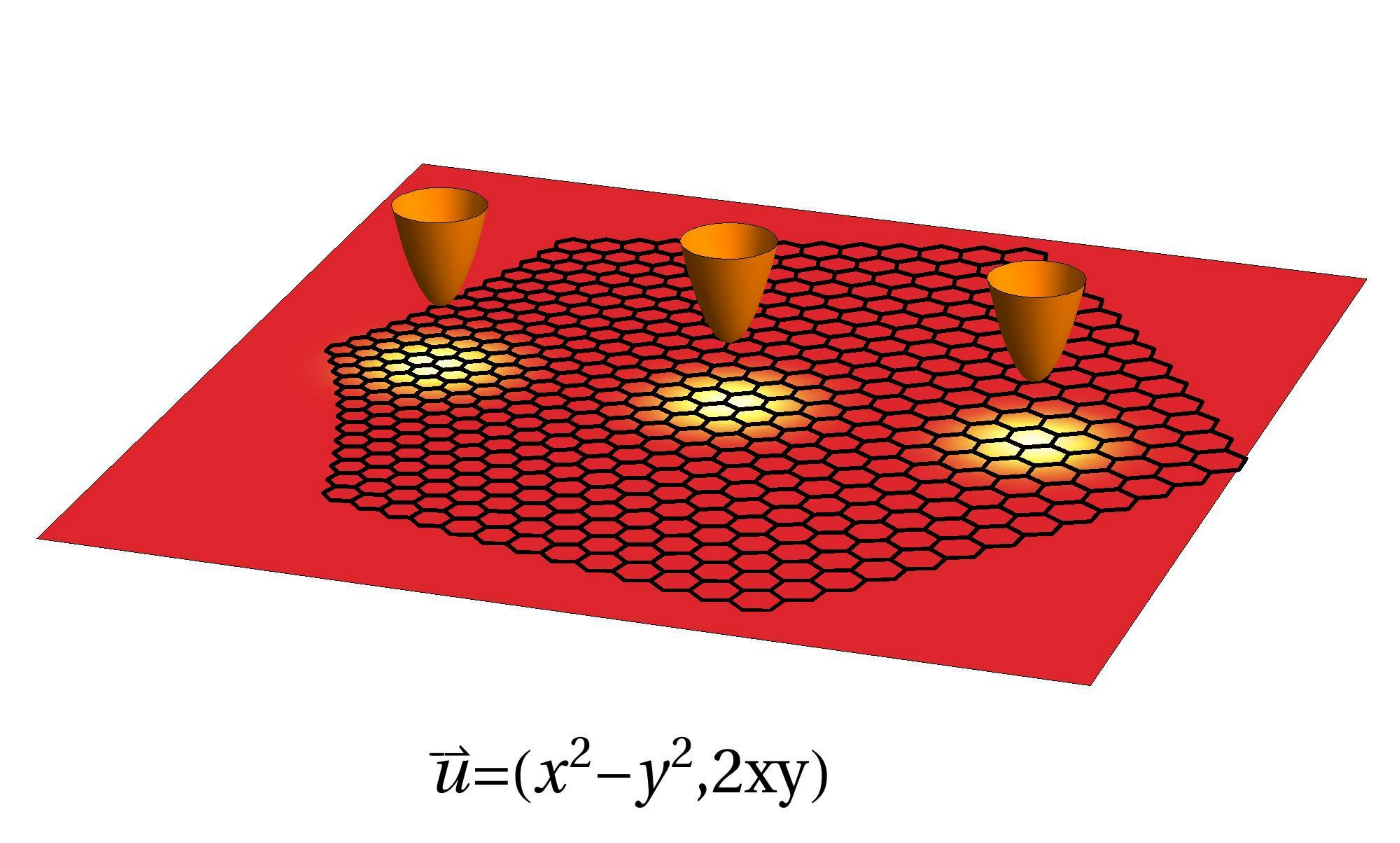}
\includegraphics[width=9cm]{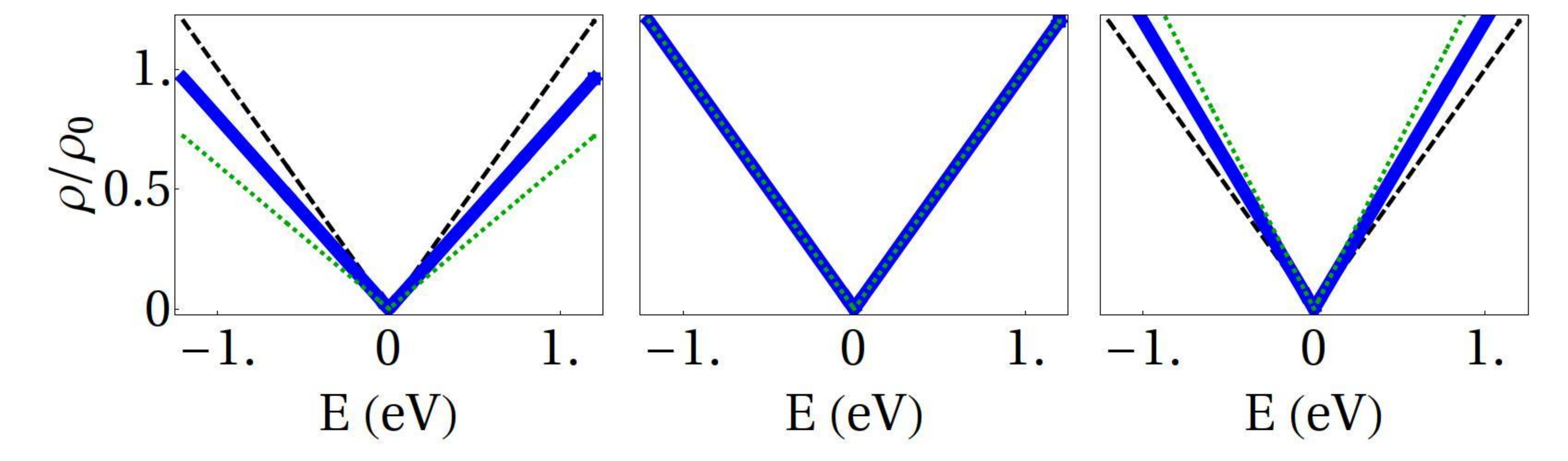}
\caption{Pictorial view of the strain field discussed in the text and the changes it produces in
the density of states. The dotted (green) line represents the contribution from $\beta$ dependent
terms alone, while the thick (blue) line represents the total correction including frame effects.
The black, dashed line represents the density of states of the perfect lattice. 
The three plots correspond to $x=-L,0,L$ for the displacement discussed in the text with
$u_{max}=0.2$.}
\label{stmfig2}
\end{center}
\end{figure}
\section{Some physical examples}
To see the physical implications of this work to actual measurements we now work out some practical
examples. Consider first a density of states measurement. The frame effects discussed are rather
trivial in this case but enough to exemplify the issue. The effect of the coordinate change will
affect STM measurements when the tip resolution is large in units of the  lattice constant (no
atomic resolution). The local density of states (LDOS) in the lab frame can be computed
approximately in the local limit, for a sufficiently smooth $u_{ij}$. To do this, $\rho(E,u_{ij})$
is computed assuming $u_{ij}$ is constant, and then its dependence on the position is restored in the
final expression $\rho(E,x) \equiv \rho(E,u_{ij}(x))$. The LDOS can be computed in momentum space
\begin{equation}
\rho (E) = \int dq_x d q_y tr (E-H(q_x,q_y))^{-1}
\end{equation}
with the Hamiltonian \eqref{main1} by diagonalizing $v_{ij}$, which amounts to a change of  integration
variables 
\begin{equation}
\rho (E) =  \int \frac{dq_+ dq_-}{v_+v_-}tr (E-H_0)^{-1},
\end{equation}
with $H_0$ the unperturbed Hamiltonian and $v_{\pm}$ the velocity eigenvalues. This yields
\begin{equation}
\rho (E) =  \frac{4}{2\pi} \frac{E}{v_+v_-} = \rho_0 (E) \frac{v_0^2}{v_+v_-}
\end{equation}
(the factor of $4$ is due spin and valley degeneracy) which to first order in strain can be computed to
give
\begin{equation}
\rho(E,x) = \rho_0(E) \left[(1+\beta\,\text{tr}\; u-\text{tr}\;\tilde u)\right] .
\end{equation}

A  simple but interesting example is provided by    in-plane strains that are quadratic in the position, such as
those associated to the triangular bumps that led to the observation of pseudo--Landau levels in STM
\cite{DWetal11} and that have been explicitly produced in artificial graphene\cite{GMetal12}.
Remember that the TB
gauge field associated to a strain tensor $u_{ij}$ is $\vec A \propto (u_{xx}-u_{yy},-2u_{xy})$.
Consider first a deformation vector given by $\vec u=(x^2-y^2,2xy)u_{max}/4L$ shown in the
upper part of
Fig.~\ref{stmfig2}. It is easy to see that the associated pseudomagnetic field will be zero in this
case.  
The trace of the strain tensor is tr u = $u_{max}x/L$,
hence a line scan along the y direction will give a perfect constant V shape ($\rho(E, x)\sim \vert
E\vert$), while along the x direction there will be a dilatation effect such that $\rho(E, x)\sim
(1+u_{max}(\beta-1)x/L)\vert
E\vert$,  as depicted in the lower part of Fig.~\ref{stmfig2} for different values of $x$. Due to the frame effects discussed in this work there is an additional, material
independent change in the magnitude of the LDOS that adds on top of the $\beta$ dependent
contributions. This is important to consider if one wants to measure the space-dependent Fermi
velocity from a local probe with resolution larger than the lattice constant. 

An interesting thing happens if we now consider the same deformation vector but exchange $u_x$ and
$u_y$, i.e.,  $\vec u \propto (2xy,x^2-y^2)$. 
In this case
there will be no volume effect (tr u=0) and the strain will give rise to a constant pseudomagnetic
field whose associated density of states will show similar Landau levels oscillations  along any
scanline. 
A $90$ degree rotation of the strain deformation 
will
give the same V shape with a Fermi velocity increasing this time along x=const. Finally, for  the
strain $\vec u \propto (x^2-y^2,-2xy)$
both the trace and the pseudo-magnetic field
will be zero and there will be no effect altogether. It can be shown that the geometric vector field
coming from the frame change does not affect the DOS at the linear order in $u_{ij}$ considered in this
work.

On the other hand, these examples are a simple illustration of the fact that the Honeycomb lattice is
very anisotropic and, of course, does not have full rotational symmetry \footnote{See for
example ref. \onlinecite{MF12} where the dependence of the pseudomagnetic field on the lattice
orientation was explored}; hence similar looking
deformations give rise to very different effects in the STM images. The important point is
that, in the case of general strain, the frame effects discussed in this work will be responsible
for additional spatial modulation of the intensity of the LDOS while preserving its energy
dependence. 

Frame effects will also be important when the absolute orientation of the lattice changes locally.
An example of this effect can be observed in the polarization dependence of ARPES signal
\cite{HPetal11}. 
The usual ARPES pictures of
Dirac cones see only one half of the cones, due to the form of the matrix element of the lattice electron at
the K point with the free electron that comes out. This effect sees the absolute orientation of the
lattice: if the lattice is rotated with respect to the polarization of light, the part of the Dirac
cone that is observed also rotates. 
As before, in order to  describe the physics in the lab frame, vectors in the
crystal frame have to be rotated to the lab frame. This is again a $\beta$-independent
contribution.  Note, however, that the suppression of part of the observed Dirac cones in ARPES is
due to the interference between photoelectrons emitted from the two sublattices and, as such, goes
beyond the  continuum limit considered in this paper.  Effects of local lattice rotations in ARPES
have been reported recently in
\onlinecite{WBetal12}. The frame effects associated to lattice rotations could also be observed in
ref.
optical experiments like those described in ref. \onlinecite{PRetal11}.
\section{Conclusions}
As a summary, we have shown that the TB description of general crystal systems  on distorted lattices
must be supplemented with geometric terms  originating in  the change of coordinates needed to describe  interactions with external probes.  These are of course always present in the experiments. The
correct Hamiltonian to use when trying to fit experiments is  $H_{lab}=H_{TB}+H_{Geom}$.
The new terms are material independent and different from the usual gauge fields arising from
deformation induced changes in the hopping parameter. We have worked out in detail the case of
strained graphene and tried to clarify some confusions  in the literature. We have seen that the
extra terms 
are of the same form as those already
present in the complete TB Hamiltonian~\eqref{HTBcomplete}, but come with $\beta$--independent coefficients.  Moreover,  aside from the well known pseudogauge fields in Eq.~\eqref{gaugefield},  the only
 vector  field in strained graphene is the  connection $\Gamma_i$ (also present in the geometric
formalism~\cite{JSV12}), which is compatible with the symmetry analysis\cite{M07,WZ10,Lin11,JMSV12b}
and required by the hermiticity of the hamiltonian whenever we have a position dependent Fermi
velocity. We have clarified that $\Gamma_i$  is not
a gauge field and will not give rise to the standard Landau levels in the density of states, 
although it may have other physical effects, such as pseudospin precession. We have also shown that
the extra gauge fields claimed in ref. \onlinecite{KPetal12} can be gauged away and do not lead to
physical consequences. 
The frame effects described in this work will be
relevant to local experiments with resolution $\lambda \gg a$, for which a continuum limit is
appropriate.

\begin{acknowledgments}
We specially thank M. Sturla for very useful conversations. Discussions with B. Amorim, A. Cortijo,
D. Faria, A. G. Grushin, F. Guinea, H. Ochoa,  A. Salas, and N. Sandler are also acknowledged.
This research was supported in part by the Spanish MECD grants FIS2008-00124, FIS2011-23713,
PIB2010BZ-00512, FPA2009-10612, the Spanish Consolider-Ingenio 2010 Programme CPAN (CSD2007- 00042)
and   by the Basque Government grant  IT559-10.  F. de J. acknowledges support  from the ``Programa
Nacional de Movilidad de Recursos Humanos" (Spanish MECD).
\end{acknowledgments}

\appendix
\section{The tight binding derivation in the lab frame.}
\label{covariantTB}
The extra terms to be added to the  standard  TB calculation due to frame effects can also be obtained  by redoing the TB
calculation directly in the lab frame. That is, we consider the TB Hamiltonian $H=-\sum_{<ij>}t_{ij}a^\dagger_i b_j+{\rm
h.c.}$,  but now we map the labels to positions in the lab frame 
\beq
H=-t\sum_{\vec y,n}a^\dagger_{\vec y}\;b_{\vec y+\vec\delta_n}+{\rm h.c.} ,
\label{hlab}
\eeq
where $\vec y=\vec x+\vec u(x)$ with    $\vec x=m_1\vec a_1+m_2 \vec a_2$, and $\vec\delta_n$  are the three nearest neighbour vectors (we follow ref. \onlinecite{JSV12} for
their definition and other conventions). As our interest here is in the $\beta$-independent terms generated by the change of frames, we have assumed that the hopping parameters $t_{ij}$ take their equilibrium value $t$.
 The real meaning  of the relabeling in~\eqref{hlab} is that non-equilibrium atomic positions are  used in the Fourier expansions
\begin{align}
a_y^\dagger&=\sum_k e^{-i\vec k\cdot[\vec x+\vec u(x)]}a^\dagger_k ,\nonumber \\
b_{y+\delta_n}&=\sum_k e^{i\vec k\cdot[\vec x+\vec \delta_n+\vec u(x+\delta_n)]}b_k .
\label{fourier}
\end{align}
 Note that, due to the fact that crystal momentum $\vec k$ is purely two-dimensional, only the in-plane components $\vec u$  of a three-dimensional
displacement $(\vec u(x), h(x))$ will appear in   the Fourier expansions in~\eqref{fourier}. As a consequence, only the linear piece $\tilde u_{ij}$  of the strain tensor can give rise to frame effects, while the out of plane contribution   $\partial_i h\partial_j h$ does not play any role in this regard. The same conclusion was reached in the main text by noting that only $\vec u$ enters the coordinate transformation that relates crystal and lab frames.

To
see how this will change the effective theory at the K-point, it is instructive to analyze $a_y$
further
before computing the Hamiltonian. If we restrict the states to $\vec k = \vec K + \vec{\delta k}$
with $\delta k<\Lambda$, we get
\beq
a_y=e^{i\vec K\cdot\vec x}e^{i\vec K\cdot\vec u(x)}\sum_{\delta k}^\Lambda e^{i\vec {\delta k}\cdot\vec x}
e^{i\vec {\delta k}\cdot\vec u(x)} a_k
\eeq
and, comparing with the corresponding expression $a_x=e^{i(\vec K+\vec \delta k)\cdot\vec x}a_k$ in the crystal frame, we observe two new contributions.
The first one is $e^{i\vec K\cdot \vec u(x)}$, which we can factor outside the integral. This is a trivial
phase factor that
can be reabsorbed into $a_y$ by a gauge transformation and has no effect on the physics.  As shown below, if
we
do not reabsorb this phase, it will show up in the effective theory as a new gauge field $A_i =
\partial_i(K_ju_j) = (\tilde{u}_{ij}+\omega\epsilon_{ij})K_j$. But this gauge field
has zero curl by construction 
and
produces no pseudo-magnetic fields, even for position dependent strains. The second term
$e^{i\vec{\delta k}\cdot\vec u(x_n)}$ cannot be eliminated by a gauge transformation in this way, and
will induce extra terms in the Hamiltonian which precisely correspond to those in Eq.~\eqref{trans} after the field rescaling~\eqref{rescaling} is performed. 

Back to the actual computation, plugging \eqref{fourier} into \eqref{hlab} gives
\beq
H=-t\sum_{x,n}\sum_{k,k'}e^{-i\vec k\cdot(\vec x+\vec u(x))}e^{i\vec k'\cdot(\vec x+\vec \delta_n+\vec u(x+ \delta_n))}
a^\dagger_kb_{k'}+h.c.
\eeq
 It is convenient to use a symmetric parametrization for the momenta: $k\to k+q/2, k'\to k-q/2$, which corresponds to the symmetric derivative convention in~\eqref{Hcrystal}. Expanding to linear order in $u$ yields
\begin{align}
&H=-t\sum_{x,n}\sum_{k,q}e^{-i\vec q\cdot\vec x} e^{i(\vec k-\vec q/2)\cdot\vec \delta_n}
a^\dagger_{k+q/2}b_{k-q/2}\times \Bigl[1+ \Bigr.\nonumber\\
&\left. -\frac{i}{2}\vec q\cdot(\vec u(x)+\vec u(x+\delta_n))-
i\vec k\cdot(\vec u(x)-\vec u(x+\delta_n))\right] +h.c.
\end{align}
which, in terms of  the Fourier coefficients of  the displacement
$u(x)=\sum_q e^{i\vec q\cdot\vec x}u(q)$ can be rewritten as
\begin{align}
&H=-t\sum_{n,k,q} e^{i(\vec k-\vec q/2)\cdot\vec \delta_n} a^\dagger_{k+q/2}b_{k-q/2} \times \nonumber \\
 &\left[\delta(\vec q)-\frac{i}{2}\vec u(q)\cdot\left(\vec q\,(1+e^{i\vec q\cdot\vec \delta_n})+
2\vec k\,(1-e^{i\vec q\cdot\vec \delta_n})\right)\right] +h.c.
\end{align}

Expanding  around the K-point and performing the sums over $n$ as usual  yields the matrix
element
\beq
H_{k,q}=\frac{3 t a }{2}[\delta(q)\sigma_i k_i +iq_iu_j(q)\sigma_i(K_j+k_j)-iq_iu_i(q)\sigma_jk_j],
\eeq
where we have redefined $\delta k\to k$. Replacing $iq_iu_j(q)=\tilde{u}_{ij}(q)+\omega(q) \epsilon_{ij}$ we finally obtain
\begin{align}
H_{k,q}&=v_0[\delta(q)\sigma_i k_i +
(\tilde{u}_{ij}+\omega \epsilon_{ij})\sigma_i K_j\non \\
 &+
(\tilde{u}_{ij}+\omega \epsilon_{ij})\sigma_ik_j-\tilde{u}_{ii}\sigma_jk_j].
\end{align}
The last two terms
are precisely those obtained from the direct coordinate transformation~\eqref{1}-\eqref{2} of the continuum Dirac
equation before the field rescaling~\eqref{rescaling}, which eliminates the term proportional to $\tilde{u}_{ii}$.
 As anticipated, there is also a seemingly new gauge field 
\beq
A_i=(\tilde{u}_{ij}+\omega \epsilon_{ij})K_j=\partial_i(u_j K_j)
\eeq
which is a total derivative and has zero associated magnetic field~\footnote{Note that the
magnetic fields shown in ref. \onlinecite{KPetal12} were computed with the full strain tensor rather
than its in-plane part, which is incorrect. The magnetic field is zero by
construction\cite{KPetal12b}.}. It can be completely 
eliminated by the  gauge transformation $\psi\to e^{-i\vec K\cdot\vec u(x)}\psi$.
\bibliography{Gauge3}

\end{document}